\documentclass[amsmath,amssymb,aps,prb,twocolumn,groupedaddress,showpacs,floatfix]{revtex4}
\usepackage{graphicx}

\begin{document}

\title{  Spin waves and local magnetizations on the Penrose tiling  }

\author{ Attila Szallas }
\author{ Anuradha Jagannathan }

\affiliation{Laboratoire de Physique des Solides, CNRS-UMR 8502, Universit\'e
Paris-Sud, 91405 Orsay, France }

\date{\today}

\begin{abstract}
We consider a Heisenberg antiferromagnet on the Penrose
tiling, a quasiperiodic system having an inhomogeneous
Neel-ordered ground state. Spin wave energies and wavefunctions are studied in the linear spin wave approximation. A linear dispersion law is found at low energies, as in other bipartite antiferromagnets, with an effective spin wave velocity lower than in the square lattice. Spatial properties of eigenmodes are characterized in several different ways. At low energies, eigenstates are relatively extended, and show multifractal scaling. At higher energies, states are more localized, and, depending on the energy, confined to sites of a specified coordination number. The ground state energy of this antiferromagnet, and local staggered magnetizations are calculated. Perpendicular space projections are presented in order to show the underlying simplicity of this ``complex" ground state.
A simple analytical model, the two-tier Heisenberg star, is presented to explain the staggered magnetization distribution in this antiferromagnetic system.
\end{abstract}


\maketitle
\section{Introduction}
Theoretical investigations of the Heisenberg
model involving localized spins on a quasiperiodic structure
are motivated by experiments performed on a class of magnetic
quasicrystals, the icosahedral ZnMgR (R: rare-earth) alloys \cite{znmgr}. Short
range antiferromagnetic correlations and anomalous slow dynamics
have been observed at temperatures below 4K in ZnMgHo,
and below 5.8 K in ZnMgTb \cite{sato}. These experiments raise the question of what
types of magnetic ordering can exist in systems with quasiperiodic structural order.
Several theoretical studies of both classical \cite{vedmed,matsuo} and
quantum \cite{wess1,wess2,ajphysrevl04,ajphysrevb05} models of interacting local
moments in two dimensional quasiperiodic geometries  have been
carried out. These calculations show ground states characterized by
a complex inhomogeneous magnetic order.
This quasiperiodic ground state, and magnon modes are investigated in detail in the present paper.

We consider spins located on
vertices of the Penrose tiling, which was introduced \cite{penref} as an example of a
determistic nonperiodic structure and later found to be relevant for
quasicrystals \cite{steinhardt}.
The Penrose tiling is
the planar version of the three
dimensional icosahedral tiling used in the description of a large number of
quasicrystalline alloys. At T=0, since
there is no frustration in the model, one expects a N\'eel ordered
ground state, with equal and oppositely directed sublattice
magnetizations. We will consider the case of $S=\frac{1}{2}$, the case of greatest theoretical interest, as
quantum effects are expected to be strongest. An
interesting related problem concerns the entanglement
properties of spins in such hierarchical environments, a question
that is currently receiving attention in view of the applications to
quantum computing. 

This paper follows an earlier Brief
Report \cite{brep2007}, where some early results were
presented. In this paper we describe the method used in some detail, and we present a more complete description
of the ground state properties, spin wave spectrum, magnon 
velocity, and eigenmodes. The numerical calculations are done within the linearized spin wave (LSW) approximation, using  periodic approximants of the Penrose tiling. We show as well that the magnetization results can be qualitatively explained in terms of a simple analytic model of tree clusters.  

In Sec.II we provide some background on the Heisenberg
antiferromagnet as well as the Penrose tiling for those not familiar with this quasiperiodic
structure. Sec.III presents the method used, and
describes the finite samples that we studied numerically. Sec.IV
presents results for the magnon spectrum and density of states. In
Sec.V we give the results for the ground state energy, and discuss the
inhomogeneous local staggered magnetizations. Sec.VI presents a simple analytical
 model to explain the numerical results. Sec.VII resumes the main
conclusions.

\section{Heisenberg Hamiltonian and the Penrose tiling}

\begin{figure}[h] \begin{center}
\includegraphics[width=8cm]{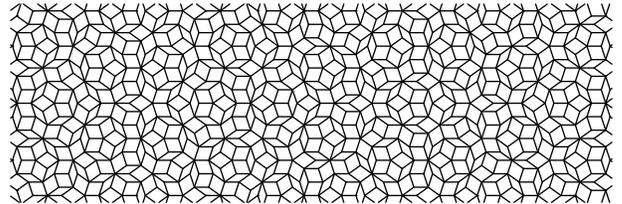}
\caption{\label{fig:tiling} A finite patch of the perfect Penrose tiling.}
\end{center}\end{figure}

\subsection{The Heisenberg Hamiltonian.}
We consider the nearest-neighbor antiferromagnetic spin-$\frac{1}{2}$ Heisenberg model,
\begin{equation}
 \label{eq:H-spin}
 H=J \sum_{\langle i,j \rangle} {\mathbf S}_i \cdot {\mathbf S}_j,\quad J>0,
\end{equation}
where $\langle i,j\rangle $ are pairs of linked vertices of the
Penrose tiling. The antiferromagnetic coupling $J$ is taken to be
the same for all bonds. The tiling is bipartite, in that sites belong to one of two
sublattices A and B, and the $J$-terms couple a spin on sublattice A 
to spins lying on sublattice B and vice-versa. This property ensures that there is no
frustration, i.e., if one considers classical spin variables, the
ground state is one for which all bonds are ``satisfied" -- with all
the A-sublattice spins pointing in one direction and B-sublattice
spins pointing in the opposite direction. This is our reference
state with long range staggered magnetic order for our calculations at $T=0$ using the spin wave approximation.
At finite temperature, the long range order will be destroyed due to the Mermin-Wagner-Hohenberg theorem \cite{mwh}, although short range correlations will persist.
We now recall some properties of the ground state
of some unfrustrated Heisenberg antiferromagnets, shown in earlier studies.

\subsubsection{Role of the dimension and the spin} When the spins are quantum variables, the energy of the
ground state is lower than the classical value, due to quantum
fluctuations or ``spin waves" in the case of periodic structures. As
studies using a variety of analytical and numerical methods have
shown (see review \cite{manousakis}), the effects of quantum
fluctuations vary depending on the dimension, on the spin quantum
number $S$, and the type of structure. Quantum
fluctuations are expected to become smaller as $S$ increases, and as the dimension
increases. It can be shown that, for hypercubic lattices, 
the spin wave expansion in fact is an expansion in $1/zS$, where $z=2d$.
For $d=2$ and $S=\frac{1}{2}$, as in the square lattice
or the honeycomb lattice, it was an open question for some time
whether quantum fluctuations would be strong enough to destroy long
range order, until enough evidence \cite{sqlatt} was presented in
favor of long range order.

\subsubsection{Role of the coordination number} In a given dimension, and for simple lattices, one can ask what the
local quantum fluctuations are when the coordination number is
changed. In two dimensions, two examples of unfrustrated
systems are the square lattice with $z=4$ and the honeycomb lattice with $z=3$
(see the review in \cite{richter}). Spin wave
calculations, as well as Quantum Monte Carlo calculations \cite{sqlatt,honey} have shown
that the order parameter is smaller
in the $z=3$ case. The
values of the staggered magnetization, defined by the value of $m_s
= \vert \langle S_i^z\rangle\vert$ as obtained by QMC calculations are  $ m_s^{sq}=0.3173$ and $m_s^{hc}= 0.2788$ (with $sq$ and $hc$
standing for square lattice and honeycomb lattice  
respectively).
This is in accord with the already
remarked tendency towards a less classical behavior for systems of
lower dimension, hence fewer nearest neighbors.

In contrast, as pointed out in \cite{ajrmsw}, when the coordination
number is not constant, the local magnetization tends to be in fact
larger when $z$ is smaller. In the dice lattice, where sites can
have $z=3$ and $z=6$, spin wave theory and QMC calculations have
shown that it is the small $z$ sites that have the larger value of
the local staggered magnetization, with $ m_s^{z=6} = 0.3754$ while
$ m_s^{z=3} = 0.4381 $. This result shows the ``counter-intuitive" 
trend towards a $less$
classical behavior for sites of bigger $z$ in structures having a
distribution of $z$ values. Other systems showing this tendency
include the quasiperiodic octagonal tiling and, as shown in
\cite{brep2007}, the Penrose
tiling.

\subsection{The Penrose tiling.}
We now describe the Penrose tiling, along with some of its
properties. As shown in Fig.\ref{fig:tiling}, which shows a finite
portion of a perfect tiling, the tiles of this structure are a pair
of rhombuses of angles $\pi/5$ (thin rhombus) and $2 \pi/5$ (thick
rhombus). The edge length will be taken to be $a=1$. 

\subsubsection{Symmetries} One remarkable
feature is that despite the absence of perfect translational
invariance, patterns of arbitrarily large size in different regions
of the tiling can be made to overlap (``local isomorphism"). The mean repetition distance of a pattern of 
linear size R is proportional to R, a property that replaces the strict
translational invariance of crystalline structures. Rotational
invariance holds in the same ``weak" sense -- for any given pattern,
its equivalent under rotation by a multiple of $2\pi/5$ can be found
elsewhere on the tiling.

The Penrose tiling possesses a
hierarchical symmetry, being invariant under so-called inflation and
deflation transformations. Inflation is a reversible operation which
can be thought of as a set of decimations of vertices of the tiling,
followed by a re-connection of the new vertices. The new tiling is
defined on a length scale that is bigger by the factor
$\tau=\frac{\sqrt{5}+1}{2}$ (the golden mean). The equivalence of
the old and new tilings means here that no environments are created
or destroyed in the process of inflation or deflation, and that one
can find an exact match between any arbitrary (finite) regions of
the two tilings.

A rhombus-based
structure such as the one shown in Fig.\ref{fig:tiling} is a
two-sublattice system. On the infinite tiling, the two sublattices are 
equivalent, as regards the frequencies of the different vertices (see subsection below). 

\subsubsection{Local environments}
\begin{figure}[h] \begin{center}
\includegraphics[width=6cm]{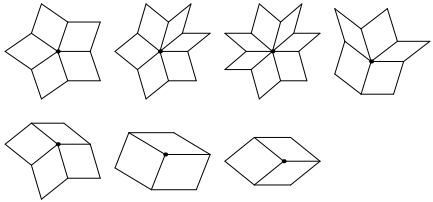}
\caption{\label{fig:starshapes} Local environments in the Penrose tiling.}
\end{center}\end{figure}

As Fig.\ref{fig:starshapes} shows, many kinds of vertices are present in
the tiling. In this paper, we have chosen to classify properties of
vertices according to their coordination number $z$. These site
coordination numbers range from 3 to 7 in the Penrose tiling, with
the average value $\overline{z}$ being exactly 4. The figure shows the seven local environments possible for 
coordination numbers $z=7,6,5,4 $ and 3. The first figure shows a five-fold symmetric site, which in fact comes in two varieties, the F (for football cluster) and the S (for star cluster). The properties of these sites under decimation are different. More will be said on these sites when we discuss the results for the ground state staggered magnetization distribution. We note that in the infinite tiling, each of the sublattices A and
B have the same distribution of
vertices of each type. 

\subsubsection{The Penrose tiling viewed as a five dimensional object}
One standard method \cite{dunaudier} to obtain the Penrose tiling is by projecting 
a subset of vertices of a five dimensional cubic lattice onto a
plane of a certain irrational orientation that is
determined by the requirement of five-fold symmetry. As described in the Appendix, 
vertices are projected only if they lie within a five-dimensional cylinder that runs parallel to the ``physical" $xy$ plane.
This approach to generating the quasiperiodic tiling allows one an
alternative visualization of the Penrose tiling in ``perpendicular
space" - i.e. the vertices are projected onto the orthogonal three
dimensional subspace. 

When the Penrose tiling is thus mapped into perpendicular space, one sees (as illustrated in the Appendix in
Fig.\ref{fig:realspace}), that the vertices now lie within four flat pentagon-shaped regions.Each family of sites of Fig.\ref{fig:starshapes} projects into a $distinct$ domain -- in other words, the perpendicular space representation allows us to separate sites according to their coordination number. 
We will use this useful property of the
Penrose tiling in order to represent the complex antiferromagnetic
ground state in a simpler way in Sec.V.

\section{The linear spin-wave approximation}
This unfrustrated Heisenberg antiferromagnet is expected to have
long range order at zero temperature, as remarked earlier. In the ground state,
spins acquire a non-zero magnetization along the z-direction, with
the total staggered magnetization of $N = \sum_i\langle \epsilon_i
S^{z}_i\rangle \neq 0$ where $\epsilon_i =\pm 1$ depends on whether
the site is on the A or B sublattice. 

Assuming that quantum
fluctuations are small, one can expand around the classical antiferromagnetic
ground state in which $\langle \epsilon_i S^{z}_i\rangle = S$ for
all sites. We introduce, according to standard definitions (see for example
\cite{manousakis}), Holstein-Primakoff boson operators for the sites
on the two sublattices: ${a_i,a^\dag_i}$ for sites on the A
sublattice, ${b_i,b^\dag_i}$ for sites on the B sublattice. The
boson operators account for deviations from the classical
configuration, with the spin deviation operator on A-sublattice
sites given by $a_i^\dag a_i = \widehat{n}_i \equiv
S-\widehat{S}_i^z$ and on B sublattice sites $b_j^\dag b_j =
\widehat{n}_j \equiv S+\widehat{S}_j^z $. After linearization, the
Hamiltonian is
\[H_{LSW} = -JS(S+1)N_b+JS\sum_{<ij>}(a_i^\dag a_i+b_j b_j^\dag+a_i^\dag b_j^\dag+b_j a_i)\]
where \(N_b = 2N \) is the number of bonds, \(N\) is the number of
sites. This quadratic Hamiltonian can be diagonalized by a
generalized  Bogoliubov transformation, as shown by White, Sparks
and Ortenburger \cite{PhysRev.139.A450}. Note that the
diagonalization can be carried out in the real space basis, and that
there is no need to introduce collective operators as in
periodic systems.

\subsection{Outline of numerical calculation scheme. }
In the expressions that follow, we number the sites such that the sites on
sublattice A come first ($i = 1, ..., N/2$), followed by the sites on sublattice
B ($i= N/2+1,...,N)$. The quadratic term in $H_{LSW}$ can then be
written in compact form in terms of the vector $X^T =
(a_1,a_2,....,a_{N/2},b^\dag_1,....,b^\dag_{N/2})$ as
\begin{eqnarray}
H_{LSW} = JS X^\dag H_2X
\end{eqnarray}
where $H_2$ is a real symmetric matrix of four $\frac{N}{2} \times
\frac{N}{2}$ blocks:
\begin{eqnarray}
H_2 = \left(\begin{array}{r|r}
Z_A & C \\
\hline
C^T & Z_B \\
\end{array} \right)
\end{eqnarray}
$Z_A$ and $Z_B$ are diagonal matrices, with $(Z_{A})_{ii}=z_i$, the
coordination numbers on sublattice A, with a similar definition on
sublattice B. The connectivites of the sites are given in the
off-diagonal blocks, C: $C_{ij}=1$ if $i$ and $j$ are nearest
neighbors and $C_{ij}=0$ otherwise. Details of the numerical
diagonalization using this real space basis are described in
\cite{wess2}, where the LSW solution for the ground state of the
octagonal tiling was obtained. The main steps of the calculation
will be outlined here for completeness.

One seeks the canonical transformation taking the set \{$a_i,b_j$ \}
to a set of boson operators \{$\alpha_m,\beta_m$ \}, in which the
Hamiltonian is diagonal:
\begin{equation}
  \label{eq:Hsw_diag}
H_{LSW} =  E_0 + \sum_{m=1}^{N/2} \Omega_m^{+} \alpha^\dag_m
\alpha_m + \Omega_m^{-} \beta^\dag_m \beta_m
\end{equation}
where $E_0$ is the ground state energy
\begin{equation}
E_0 = -JS(S+1)N_b +JS \sum_{m=1}^{N/2}\Omega_m^{-}
\label{gse.eq}
\end{equation}

Two of the eigenmodes, $a_{N/2}$ and $b_{N/2}$, correspond to zero eigenmodes, and arise due
to a global rotation of all the spins. 
These should be excluded from the diagonalization procedure, which
transforms the remaining (N-2) operators. The generalized  Bogoliubov transformation from the set $\{a,b\}$ to the
set $\{\alpha,\beta\}$ is $X_m = T_{mi}X{'}_i$ where
$X{'}^T=(\alpha_1,,....,\alpha_{N/2-1},\beta^\dag_1,....,\beta^\dag_{N/2-1})$.
 The rectangular T-matrix has the following structure:
\begin{eqnarray}
T = \left( \begin{array}{r r r r r r }
A_{1,1} & . &.& . &. & A_{1,N} \\
. & .& .& .& . &. \\
. & .& .& .& . &. \\
A_{\frac{N}{2}-1,1} & .& . &. &. & A_{\frac{N}{2}-1,N} \\
B_{1,1} & .& . &. &. & B_{1,N} \\
. & .& .& .& . &. \\
. & .& .& . & . &.\\
B_{\frac{N}{2}-1,1} & .& . &. &. & B_{\frac{N}{2}-1,N} \\
\end{array} \right)
\end{eqnarray}
The transpose of the row vectors of this matrix, denoted by $A_m$ and $B_m$, 
correspond to solutions of the eigenvalue equation
\begin{eqnarray}
gH_2 A_m &=& \Omega_m^{+}A_m \nonumber \\
gH_2 B_m &=& -\Omega_m^{-}B_m
\end{eqnarray}
where the first equation holds for matrix elements of the rows in
the upper half of $T$ (the ``positive subspace"), and the second for the rows in the lower half
of $T$ (the ``negative subspace").  $g$ is the matrix of commutators $g_{ij} = [X_i,X_j^\dag]$,
\begin{eqnarray}
g_{ij} = \pm\delta_{ij}
\end{eqnarray}
where the sign depends on whether $i$ corresponds to the positive or the negative subspace. 
The new operators X' obey the same type of commutation relations 
$[X{'}_i,X{'}_j^\dag] = g{'}_{ij} = \pm\delta_{ij}$ (where $i,j = 1,....,N/2 -1$) .

The T-matrix should satisfy 
\begin{eqnarray}
T^\dag gT = g{'} 
\label{tmatrix}\end{eqnarray}
We note that in degenerate subspaces, a generalization of the Gram-Schmidt orthogonalization must be carried out with respect to the matrix $g$, in order to satisfy Eq.\ref{tmatrix}. 

\subsection{Finite size samples for numerical calculations.}
Our numerical solution of the linearized spin wave Hamiltonian is
carried out on finite samples, using LAPACK \cite{lapack}. We have mentioned in the previous
section that quasiperiodic tilings can be considered as projections
of a simpler higher dimensional periodic systems. For our calculations, we have
considered two types of samples. The first type are finite samples obtained by projection from 5D, for
which we assumed open boundary condition. Such samples have 
the usual problem of spurious boundary modes. When considering averages over the entire spectrum, however, as one does in the computation of the order parameter, we found good agreement between these finite samples, and periodic approximants described below.  More precisely, spins deep in the interior of these samples behave in the same way as the spins in the boundary-free
 periodic approximants, described next. 

In order to apply periodic boundary conditions, we have considered the so-called
Taylor approximants of the Penrose tiling \cite{dunaudier}. These rectangular samples are obtained
by slightly tilting the projection (or applying a ``shear"), so that
the resulting projection has a finite periodic length in the x- and
y-directions, as shown in Fig.\ref{fig:taylor3}. These approximants clearly break
the hierarchical and five-fold rotational symmetry of the Penrose tiling
at length scales comparable to the repetition length. However, one can generate
Taylor approximants of bigger and bigger size, and thus approach the
infinite Penrose tiling.  Our samples have the
additional property of having an equal number of sites in the A and
B sublattices, thus ensuring a total spin of zero in the ground
state. We considered approximants of $94$, $246$, $644$,
 $1686$, $4414$ and $11556$ sites.

\section{Magnon spectrum and wavefunctions}
\subsection{Energy spectrum and density of states. Linear dispersion law.}
 The numerical Bogoliubov transformation gives the energies
$\Omega_m^{\pm}$ as well as the $T$-matrix for each of the systems
considered. The two sets of energies $\Omega_m^{+}$ and $\Omega_m^-$
become identical in the limit of infinite size when sublattices A
and B become strictly equivalent. In our case, since the two
sublattices are not identical, the numerical values are slightly
different (less than a percent for the larger systems and not
visible on the scale of our figures).

\begin{figure}[t] \begin{center}
\includegraphics[width=5cm]{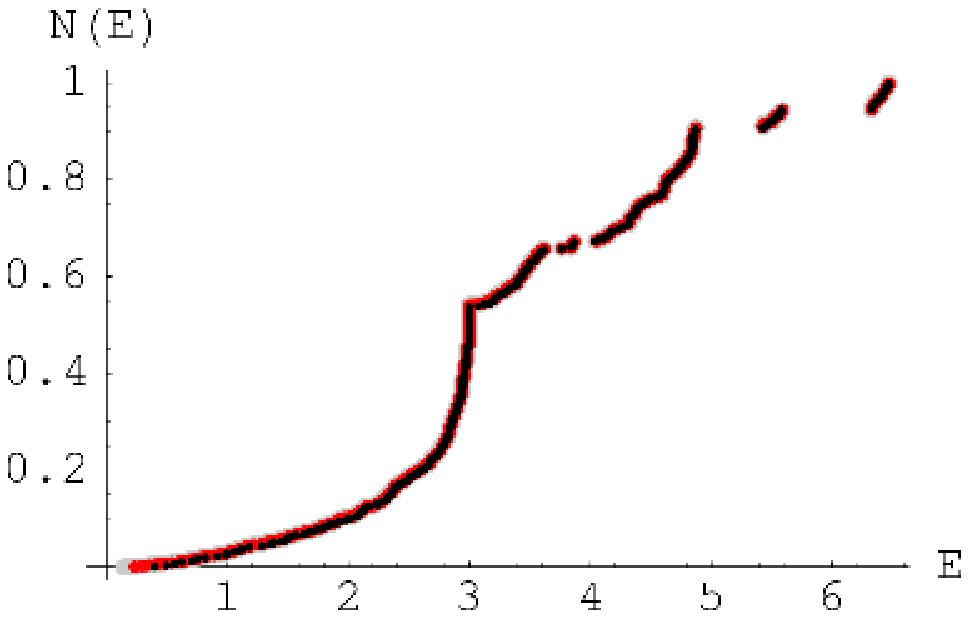}
\includegraphics[width=5cm]{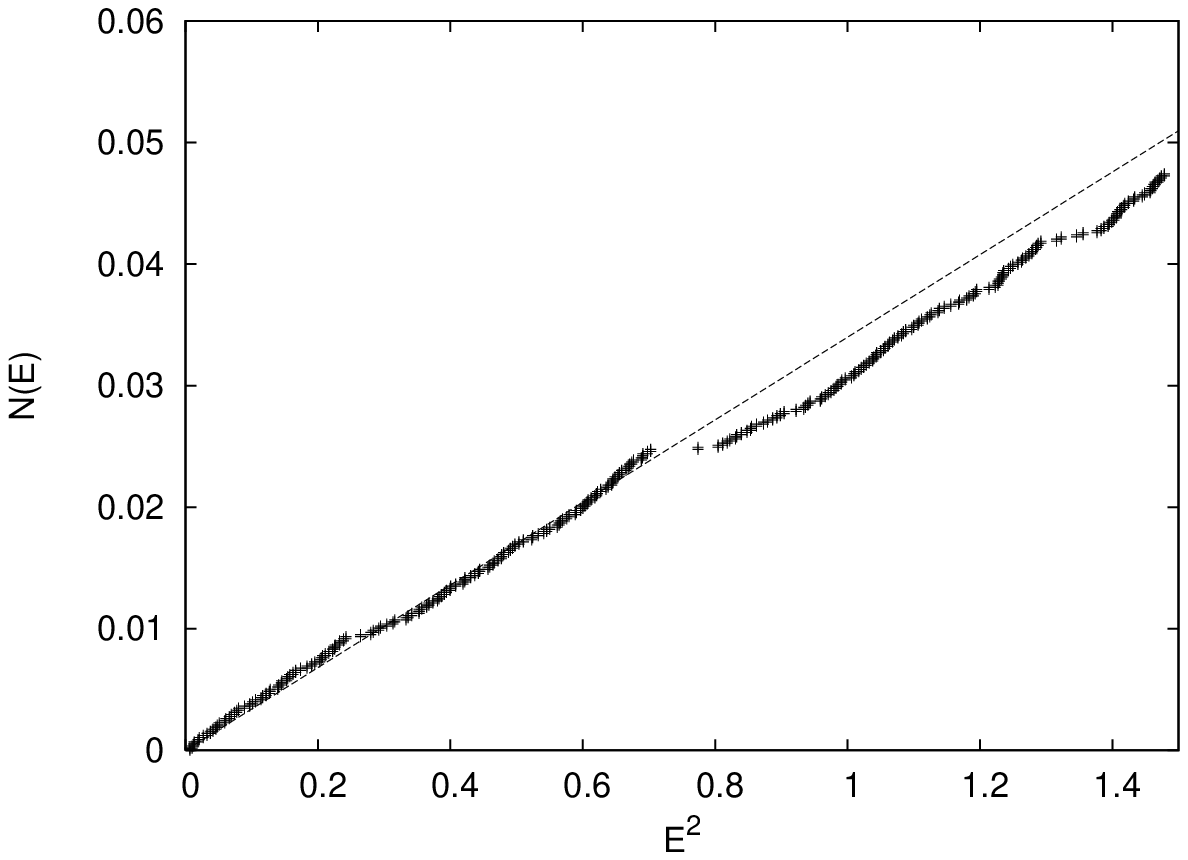}
\caption{\label{fig:IDOS_penrose} a)(Color online) Integrated density of states $N(E)$ versus $E$ (expressed in units of $J$) calculated
for three consecutive approximants ( black dots : $N=644$, red dots : $N=1686$, grey : $N=4414$) . b) Low energy tail of $N(E)$ versus $E^2$ (the straight line is a fit to the data). }
\end{center}\end{figure}

Fig.\ref{fig:IDOS_penrose}a) shows the integrated density of states (IDOS). We used the following expression for the IDOS,

\begin{equation}
N(E) = \frac{2}{N}\sum_{m=1}^{N/2} \theta (E-\Omega_m^{+})
\end{equation}
The energy $E$ is expressed in units of $J$, and the factor 2 arises because the sum includes only half of the complete spectrum. Data for for three successive Taylor approximants has been plotted, with the points in
light gray correspond to the largest system size, points in red to
the medium size, and points in black to the smallest size. The integrated DOS is only slightly size dependent, as seen by the overlapping of the three sets of data points. The salient features of the figure are the
several groups of closely spaced energy levels, the main
gaps, which are stable with increasing system size, and a discrete jump at
the energy $E=3$, corresponding to a macroscopic degeneracy, as discussed below. 

\begin{figure}[t] \begin{center}
\includegraphics[width=7cm]{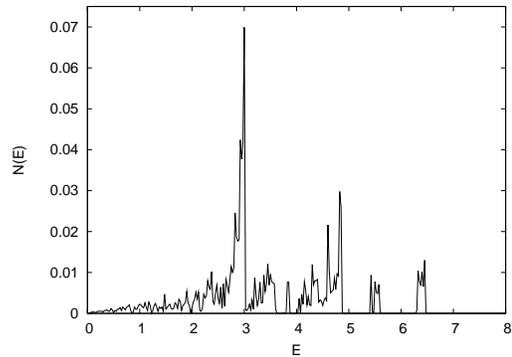}
\caption{\label{fig:DOS_penrose} Density of states calculated
for the Taylor 6 approximant (N=11556 sites). }
\end{center}\end{figure}

The locations of gaps are more easily seen in the graph of the smoothed derivative of $N(E)$ -- the density of states (DOS), which is shown in Fig.\ref{fig:DOS_penrose}. 
One sees
the characteristic fluctuating form of the density of states typical of
quasiperiodic systems.

One can distinguish several groups of energies separated by gaps.
The highest energy bands around the values $E\approx 5.4$ and
$E\approx 6.4$, correspond to wavefunctions that are localized on
the 6- and 7-fold sites, as will be shown shortly. Lying below these
in energy are the states centered primarily on the 5-fold sites. The
peak at the exact value of  $E=3$ corresponds to string-like states
living on the $z=3$ sites. The lowest energies correspond to states
of relatively extended character, as we will discuss below.

As shown in Fig.\ref{fig:IDOS_penrose}b) the low energy part of the IDOS
can be fitted to a power law of the energy, $N(E) \sim E^2$
reflecting a linear dispersion of the magnon modes in this region of
the spectrum. Fitting to a form $N(E) = E^2/(8 \pi c^2)$ gives a
sound velocity on the Penrose tiling of $c = 1.08 J$. It is interesting to compare this result for $c$ with the corresponding values
on the square lattice and the octagonal tiling, both having in common with the Penrose tiling the same
value of the classical energy per spin $E_{cl}/N = -2JS^2$. This value is
$c_{sq} =  2\sqrt{2}JS a \approx
1.41 J$ on the square lattice (for edge length $a=1$ and $S=1/2$).
On the octagonal tiling, our estimated value is $c_{octa} \approx 1.3 J$, in agreement with Milat and Wessel's IDOS data \cite{wess2}.
To resume,
an acoustic-type dispersion relation is obeyed at long
wavelengths, with a spin wave velocity which is smaller than in the octagonal 
tiling, which is in turn smaller than the value on the square lattice. 
This is presumably due to the fact that the density of sites 
is largest in the Penrose systems, followed by the octagonal tiling, and finally
the square lattice. 

\subsection{Wavefunctions.}
\begin{figure}[t] \begin{center}
\includegraphics[width=5cm]{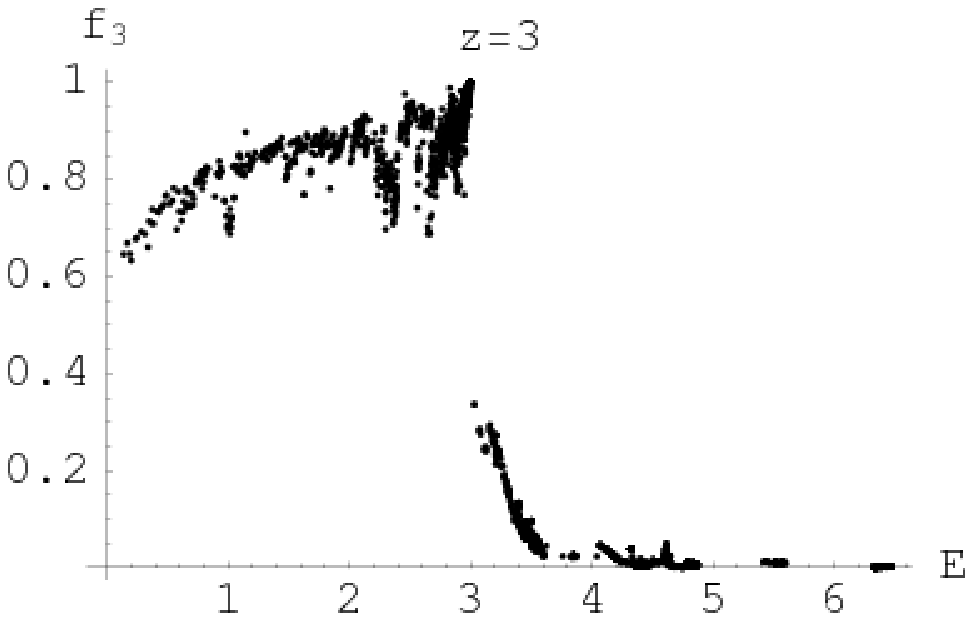}
\includegraphics[width=5cm]{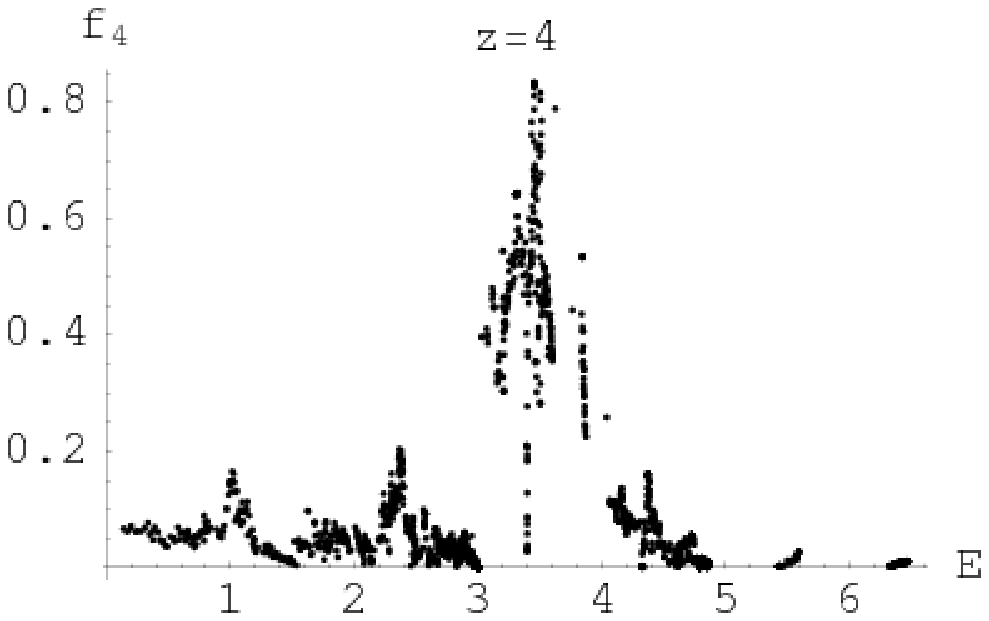}
\includegraphics[width=5cm]{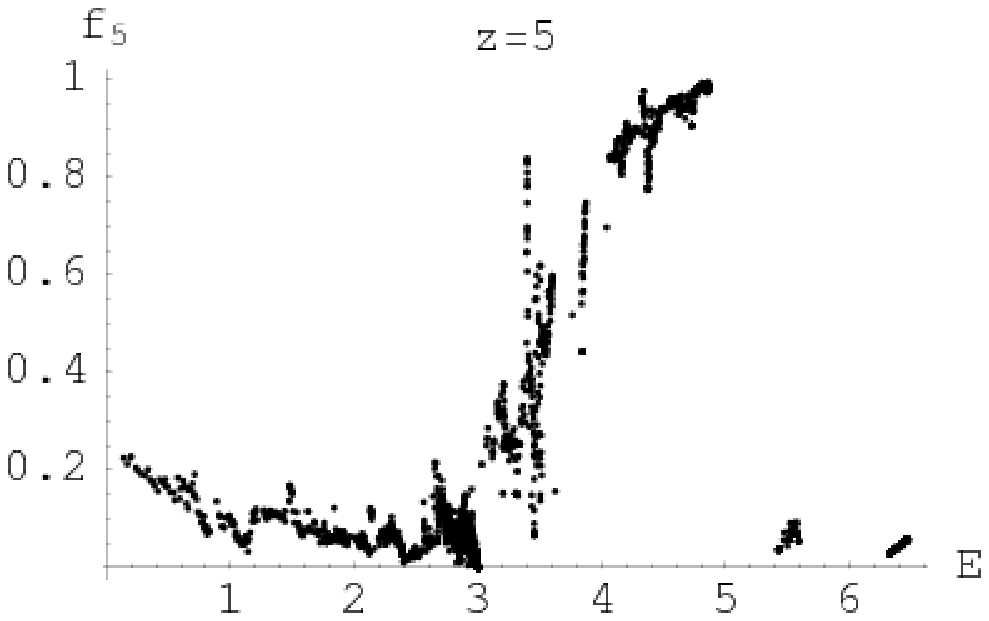}
\includegraphics[width=5cm]{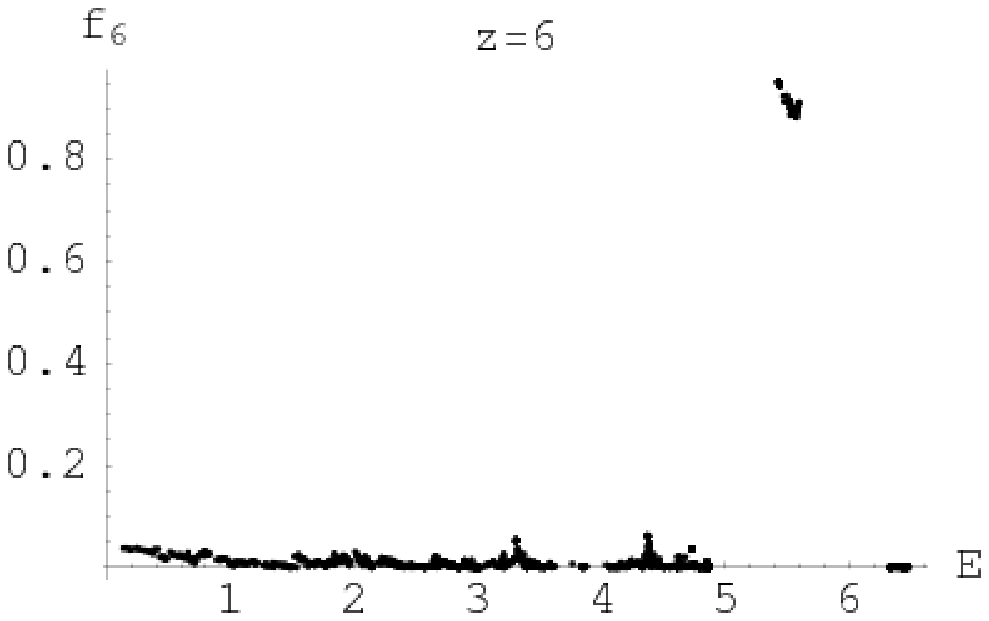}
\includegraphics[width=5cm]{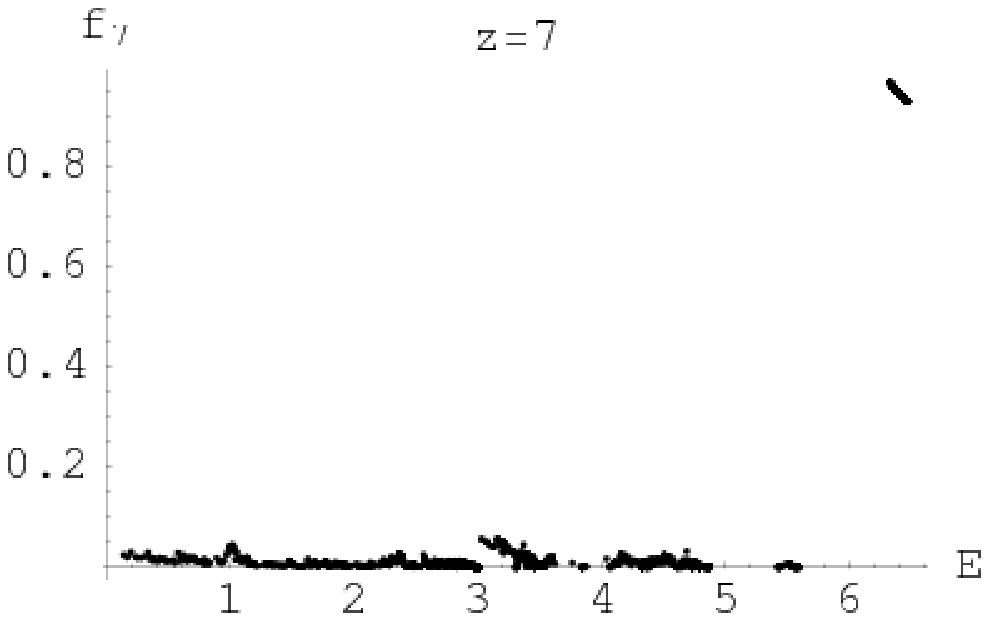}
\caption{\label{fig:partialdosfig} a) - e) Weight fractions $f_n$ (see text) as a function of $E$ for $z=n$  ($n=3,...,7$), as computed for the Taylor 5 approximant (N=4414 sites).}
\end{center}\end{figure}

We now discuss the magnon wavefunctions and their spatial characteristics for 
the different parts of the energy spectrum.
We
first show that the coordination number plays an important role in
determining the extent to which a site participates in the
wavefunction for a given eigenmode, $\psi^{(E)}$.  This can be seen
from Figs.\ref{fig:partialdosfig}a) through e), which show the weight
fractions as a function of the energy, for each of the five values of the coordination number $z$. The
weight fractions were defined as
\begin{eqnarray}
f_n = \sum_{i \in F_n} \vert \psi^{(E)}_i\vert ^2/ \sum_j \vert
\psi^{(E)}_j\vert ^2
\end{eqnarray}
where the $F_3,F_4,...$ are the subsets of sites whose coordination
numbers are $z=3,4...$. Only sites of sublattice A are considered (a similar calculation for sublattice B gives the same results).

\begin{figure}[t] \begin{center}
\includegraphics[width=8cm]{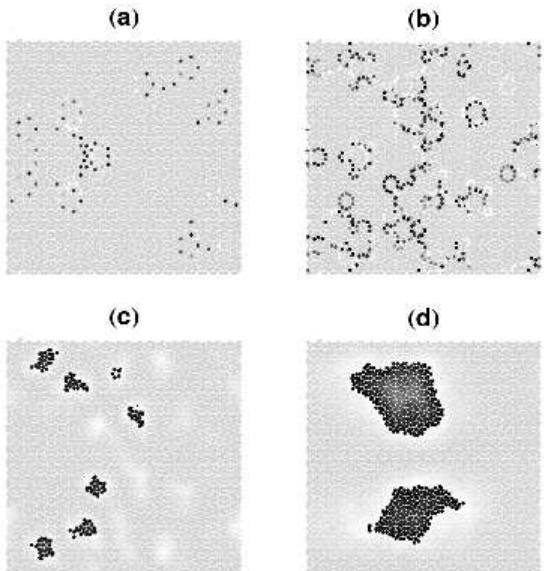}
\caption{\label{fig:wfnspar} Intensity plots representing the
probability $\psi_i(E)^2$ for different energies E (a darker shade corresponds to a higher probability): a) $E=6.469$ b)
$E=3.0000$ c) $E=0.8887$ d) $E=0.1692$. The solutions have been
obtained for the Taylor 5 approximant (N=4414 sites). }
\end{center}\end{figure}

The plots of the weight fractions show a number of interesting
properties of the magnon modes. In particular, they show the
preponderance of certain types of site in the different energy
bands. Thus it is clear from Fig.\ref{fig:partialdosfig} e)
that the highest energies
correspond to states having a large amplitude on sites of $z=7$. The highest energy band has a width of about
0.16 and is centred around $E=6.4$. Fig.\ref{fig:wfnspar}a)
represents one such state in real space, with sites are shown with
varying intensity depending on the square of the wavefunction
amplitude. The darkest spots are those on sites with $z=7$.

The group of states next highest in energy are those involving $z=6$
sites, and correspond to energies in the range $5.43 < E < 5.49$.
These states have a smaller dispersion than the states in the
topmost group of energies, reflecting the fact
that the six-fold sites are significantly fewer in number than the
seven-fold sites.

\begin{figure}[t] \begin{center}
\includegraphics[width=6cm]{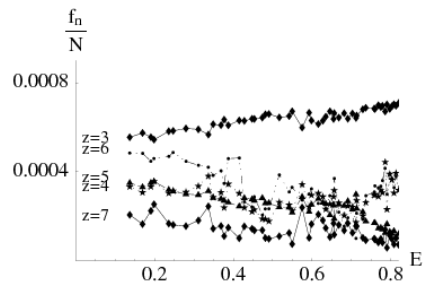}
\caption{\label{fig:allfig}  The average value of $\psi_i(E)^2$ per site for each $z$ plotted against $E$ for states in the bottom of the spectrum.}
\end{center}\end{figure}

The middle band of energies, $3 <E <5$ arises for states involving
sites of $z=5$, which account for about 30\% of the sites. From
Fig.\ref{fig:partialdosfig}c) one sees that the four-fold sites are
particularly important in a narrow range of energy within this band.

As seen by the large step in $N(E)$ at $E=3$, a large number of
degenerate states occur at this energy. This is due to wavefunctions
that have their entire support on the $z=3$ sites. These are string-like
states forming closed loops -- on length scales that range from a small
ring (around the footballs, for example) to being as large as the
system size. A linear combination of such degenerate states is shown in 
Fig.\ref{fig:wfnspar}b).

The lowest energy states , for $E \leq 1$, are the
closest to extended states. The wavefunction amplitudes depend less sensitively on the site coordination number. The fact that all sites participate is best seen from Fig.\ref{fig:allfig},
where we have plotted for each of the five $z$-values the average
probability for a given site, $f_n/N_n$, (where $N_n$ is the number of sites of the $n$th family),
as a function of the energy $E$. Clearly, for the low energy states, the probability amplitude is nonzero for all the values of $z$. On the other hand, not all of the sites participate in a given wavefunction, and for a given energy the wavefunction is mainly confined to a set of disconnected patches. The
lower the energy, the larger the patches where the wavefunction is
non-zero. Figs.\ref{fig:wfnspar}c) and d) show the states
corresponding to two energies. For $E=0.8887$ there are many closely spaced small
patches. As the energy gets smaller, the patches of non-zero amplitude get larger, along with the spacing between them,
and at
energy $E=0.1692$, for example, the wave function has
two large patches (which are of course repeated periodically, due to the
boundary conditions). 

To resume, the dimension of the support of eigenstates decreases as a function of the energy, 
from two at small $E$, to one at $E=3$, tending to zero at the highest energies.

\subsection{Perpendicular space representation of the wavefunctions}
In the preceding section we alluded to the fact that the
perpendicular space representation of the Penrose tiling is a useful
way to see the environment-dependence of spatially varying
quantities. We now illustrate this in the case of the four
wavefunctions shown in Figs.\ref{fig:wfnspar} represented in
perpendicular space by Figs.\ref{fig:wfnsperp}. Each vertex of the
Taylor approximant is mapped (see the Appendix for more details) onto a point $\{x_\perp,y_\perp, z_\perp\}$. 
We show the projection in the plane $z_\perp=2$ of the 
Penrose tiling, with regions shaded according to the local value of
the wavefunction. Specifically, the intensity of the spot at site $i$ is
proportional to the value of $\vert\psi_i(E)\vert^2$.
Fig.\ref{fig:wfnsperp}a) shows the perpendicular space projection
for the wavefunction corresponding to the energy $E=6.469$. The
spots of maximum intensity are in the region that corresponds to
$z=7$ (as seen in Fig.\ref{fig:realspace}). Fig.\ref{fig:wfnsperp}b)
shows, similarly, that the wavefunction for $E=3$ is non-zero for
the region corresponding to $z=3$. The last two figures show
wavefunctions that are delocalized in perpendicular space (ie, all
sites are involved, regardless of the value of the coordination
number).

\begin{figure}[t] \begin{center}
\includegraphics[width=8cm]{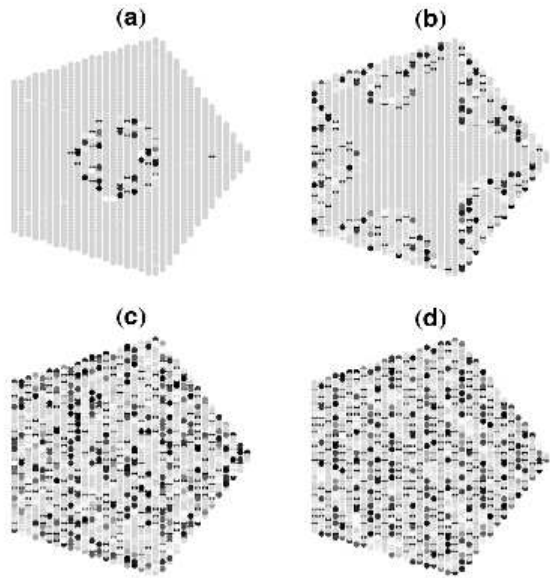}
\caption{\label{fig:wfnsperp} Intensity plots in perpendicular space representing the probability $\psi_i(E)^2$ for different energies E (a darker shade corresponds to a higher probability): a) $E=6.469$ b) $E=3.0000$ c) $E=0.8887$ d) $E=0.1692$  }
\end{center}\end{figure}

\subsection{Participation ratio and multifractal analysis.}

We present the results for the inverse participation ratio (IPR), defined by

$$ P^{-1}(E) = \frac{\sum_j \vert \psi_j(E)\vert^4}{(\sum_j \vert \psi_j(E)\vert^2)^2} $$

\begin{figure}[t] \begin{center}
\includegraphics[width=9cm]{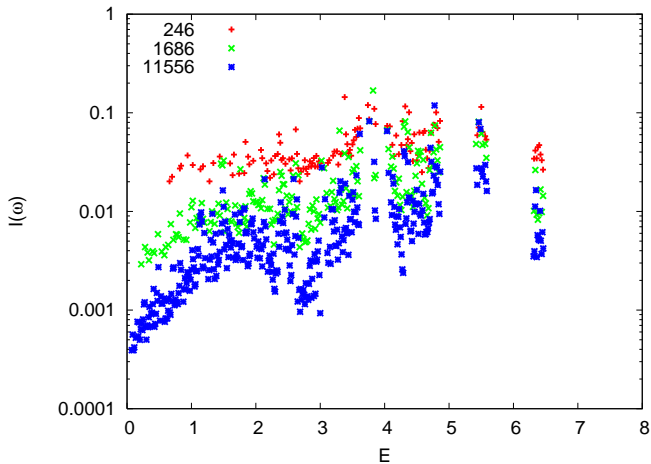}
\caption{\label{fig:ipr} Inverse participation ratio (IPR) plotted on a log scale as a function of the energy for three successive Taylor approximants }
\end{center}\end{figure}

Recall that as $N$ is increased, the inverse participation ratio decreases as $1/N$ for truly extended states, tends to a constant for localized states, and has an intermediate behavior scaling as $N^{-\beta}$ for the so-called critical states.
This quantity has been much studied in the case of tight-binding models for electrons in disordered systems, in particular, close to or at the critical disorder for the metal-insulator transition (MIT) \cite{kramer}. For electrons in the Penrose and Ammann-Kramer-Neri tilings, Grimm et al \cite{grimm} have found a distribution of values of $\beta$ ranging between 0.5 and 1.
Turning now to our spin problem, Fig.\ref{fig:ipr} shows the results for $\log (P^{-1}(E))$ versus $E$
for the Taylor approximants for three different sizes. (Note: the values have been calculated separately for each sublattice, according to the corresponding sector $\Omega^\pm$ of the eigenvalue spectrum). 
The fluctuations in the IPR tend to be quite large from one energy to the next, (note that the figure is plotted on the logarithmic scale), however, the smoothed IPR is an increasing function of $E$ over most of the energy spectrum. There is a noticeable dip in some of the IPR values as the energy approaches the value $E \approx 3$, and at $E=3$ the value of the IPR does not reflect the spatial extent of the eigenstates, because of the mixing of the macroscopically degenerate states at this energy.

Fig.\ref{fig:ipr} shows the most marked size dependence in the region of small energies, in accord with our earlier observation that states relatively delocalized at low energy. At high energies, the states are more localized, and the size dependence is accordingly smaller. 

As a general remark, results for the IPR should be treated with precaution, in the case of quasiperiodic systems, as compared with disordered ones. In the latter case, away from the MIT, the decay of the wavefunctions is typically exponential, and degenerate states are unlikely. In the quasicrystal, exactly degenerate states do occur at special values of the energy ($E=3$ here), which leads to an IPR value equal to that of extended states, due to linear combinations of localized states. Our results for the IPR, taken together with the preceding analyses of the wavefunctions, indicate that wavefunctions can be considered to be two dimensional, and power law extended in the lower end of the spectrum. The average value of $\beta$ , found by fitting the form $P^{-1}(E) \propto N^{-\beta}$ found in this region is $\beta \approx 0.9 \pm 0.1 $.

\begin{figure}[t] \begin{center}
\includegraphics[width=8cm]{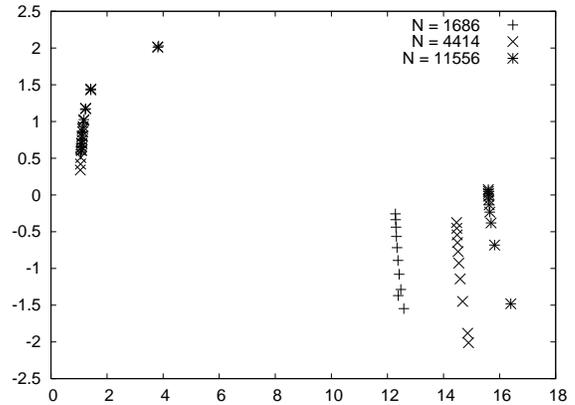}
\caption{\label{fig:multifrac} Some examples of $f(\alpha)$ plots for small energy eigenstates in three Taylor approximants. }
\end{center}\end{figure}

The multifractal character of wavefunctions for quasiperiodic Hamiltonians can be considered to be established in a number of one dimensional models \cite{kohmoto} and strongly indicated in a number of two dimensional systems including the tight-binding model for electrons \cite{grimm} and the phonon problem \cite{los3}, and even in three dimensional models \cite{los2}. The scaling properties of the wavefunctions can be investigated by calculating the $f(\alpha)$ spectrum. As $f(\alpha)$ depends on system size, and our system sizes are rather small, an extrapolation to infinite size is not without risk. Nevertheless, we have investigated the multifractal scaling properties in the low energy end of the spectrum, where the states are extended, and have an patchy structure, with peaks and valleys spaced farther and farther apart as energy decreases -- see Figs. \ref{fig:wfnspar}c) and d). We find evidence for a nontrivial distribution of exponents in the limit of large system sizes.  Fig.\ref{fig:multifrac} shows the results of a multifractal analysis of two representative low energy states for three system sizes. The shape is a priori size dependent, and state dependent as well. One can nevertheless distinguish a smooth functional form of the $f(\alpha)$ function in each case. The curves have a maximum at $\alpha(0)\sim 4$ with  $f[\alpha(0)]=2.0$, which is the dimension of the support -- also called similarity dimension -- of the wavefunction of this system. The so-called information dimension for these wavefunctions 
is given by $f[\alpha(1)] = \alpha_1$ (also known as $D_1$). It corresponds to the intersection of the $f(\alpha)$ curve and a straight line of slope 1 and we find  $\alpha_1\approx 1.5$. For comparison, for a three dimensional system at the metal-insulator transition (MIT) the critical values have been found to be \cite{grussbach} $\alpha(0)=4$ and $\alpha(1)= 2$.

\section{Results for ground state energy and staggered magnetizations}

\subsection{Ground state energy}
\begin{figure}[t] \begin{center}
\includegraphics[width=6cm]{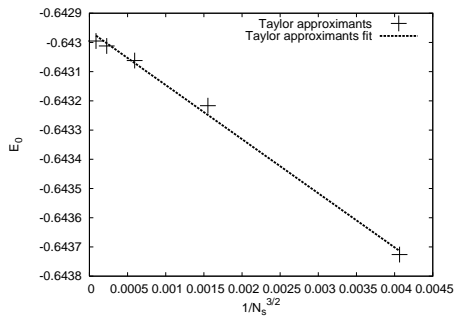}
\caption{\label{fig:GSE_penrose} The scaling of ground state energy of the Taylor approximants with system size}
\end{center}\end{figure}

The ground state energy per site is found from Eq.\ref{gse.eq} for
the periodic approximants of size N ranging from 96 to 11556. The
energies are shown plotted against $N^{-3/2}$ in the figure
\ref{fig:GSE_penrose}. This is the expected power law for the ground state
finite size correction in two dimensional periodic systems. We
find that it is
obeyed on the average in the Taylor approximants, with deviations that get smaller as size increases. 
Our extrapolation to infinite
size gives an asymptotic value $E_0 = -0.643(0) \pm 0.0001$.
The sign of the deviations varies with system size, and we believe this may be due to the
defects present in the Taylor
approximants. Since each approximant was obtained by a different
rational section in five dimensional space, the distribution of
local environments differ from sample to sample. In
\cite{commentprlaj} it was argued that the ground state energy of
the tiling may depend in a simple way on moments of the variable
$zS$. The first moment is just the average value of $zS = 4S$ in all
cases. However $\langle (zS)^{-1}\rangle$ varies from sample to
sample, and this possibly accounts for the sign and magnitude of the observed deviations from
the straight line power law behavior. 

\subsection{Staggered local magnetizations}
\begin{figure}[t] \begin{center}
\includegraphics[width=5cm]{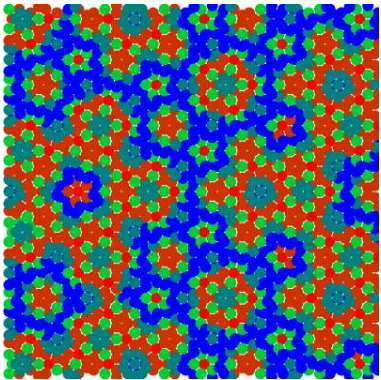}
\caption{\label{fig:realspacefig} A portion of a Taylor approximant with vertices colored according to the value of the onsite
magnetization (magnetization values : red (small), green (intermediate) blue (highest))}
\end{center}\end{figure}

The absolute value of the local magnetizations in the ground state are given in linear spin wave theory by
\begin{eqnarray}
 \label{eq:lsm}
 m_s(i)=|\langle S^z_i \rangle |=S- \sum_{k>\frac{N}{2}-1} |A_{ik}|^2 \qquad (i \leq N/2)\nonumber \\
 m_s(j)= |\langle S^z_j \rangle |=S- \sum_{k\leq\frac{N}{2}-1} |B_{jk}|^2 \qquad (j> N/2)
\end{eqnarray}
Fig.\ref{fig:realspacefig} represents how local magnetizations vary
in space on a portion of the Penrose tiling. The color of the
circles around each vertex varies from red (small magnetization) to
blue (high magnetization). The lowest values of staggered magnetization are found on a certain
subset of $z=5$ sites, as will be discussed further below.
The largest values are found on the low
coordination sites of $z=3$.

\begin{figure}[t] \begin{center}
\includegraphics[width=8cm]{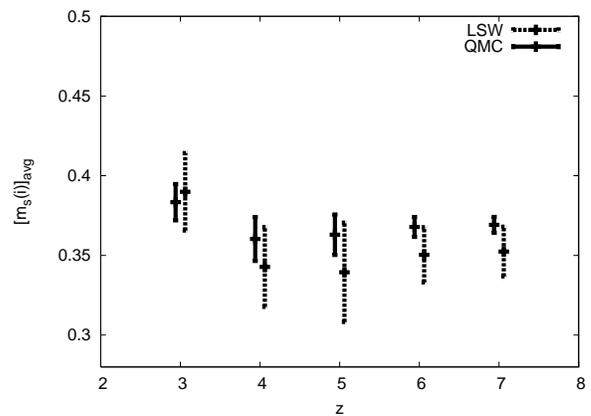}
\caption{\label{fig:mzfunc} Averages and standard deviations of the
local staggered magnetization as a function of $z$ for N=4414 ( LSW theory (dashed line) QMC (continuous line))
}
\end{center}\end{figure}

\begin{figure}[t] \begin{center}
\includegraphics[width=8cm]{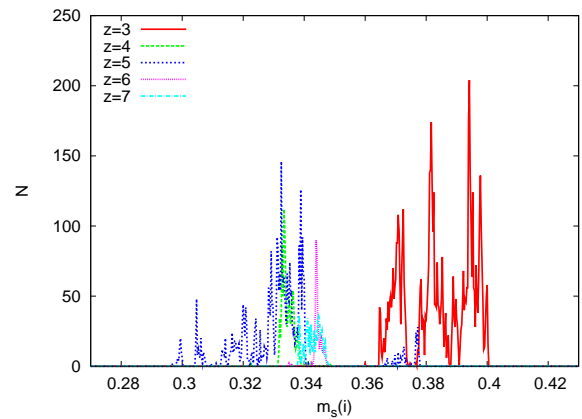}
\caption{\label{fig:distfn} Probability distribution of the local staggered magnetizations for the five different values of $z$ (N=11556)
}
\end{center}\end{figure}

Fig.\ref{fig:mzfunc} indicates how values of the local magnetizations
vary with $z$ value for system size $N=4414$. The figure shows the
average values and the standard deviations of $m_i(z)$ for each
value of $z$, as obtained in LSW theory and from QMC ,
as reported in \cite{brep2007}. Finite Penrose samples
gave the same results as the periodic
approximants when the sites on the free boundary layer were excluded
from the analysis.  The figure shows that Quantum Monte Carlo gives
a narrower spread of the magnetization. Linear spin wave theory clearly overestimates the fluctuations from average
behavior, giving a too high value for $Z=3$ and too low values for higher values of $z$. 
We show as an example, the details of the
distribution of values obtained for $z=3$ in Fig.\ref{fig:distfn}.
There is a continuum of values, as expected for a quasiperiodic
structure, but also some pronounced peaks, corresponding to specific
local configurations. We explain these features in terms of the next
nearest neighbor configurations present on the tiling, as seen in the next section.  The five-fold
sites come also in three main categories, as shown in Table 1, and
this will be taken up in more detail further below.

\begin{figure}[t] \begin{center}
\includegraphics[width=8cm]{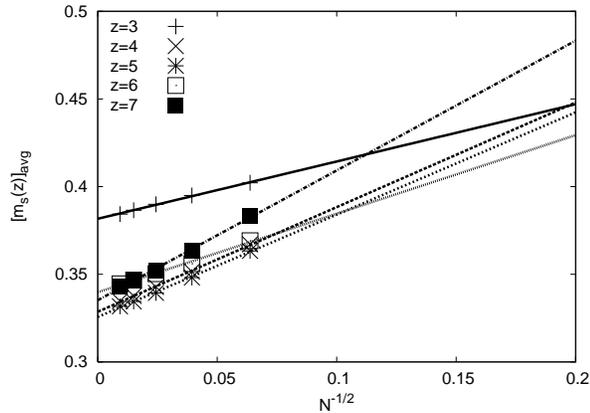}
\caption{\label{fig:avgm} System size dependence of the average local staggered magnetization for each coordination number $z$ (lines indicate fit to power law $N^{-1/2}$)}
\end{center}\end{figure}

The size dependence of the magnetizations can be seen in Fig.\ref{fig:avgm}, where we show the values of the staggered magnetization in six different sample sizes ($N=96$ to $N=11556$). We show the scaling of the average value of $m_i$ for a given $z$ calculated for each of the six approximants in Fig.\ref{fig:avgm}. The averages are seen to obey a scaling with $N^{-1/2}$ as in periodic systems. 


We now turn to the explanation of the fluctuations observed in Fig.\ref{fig:mzfunc} of the
average staggered magnetization with $z$. This
behavior can be explained with the help of a simple star cluster model,
discussed in Sec.VI.

\subsection{Perpendicular space representation of the ground state staggered magnetization}

\begin{figure}[t]
\begin{center}
\includegraphics[scale=0.4]{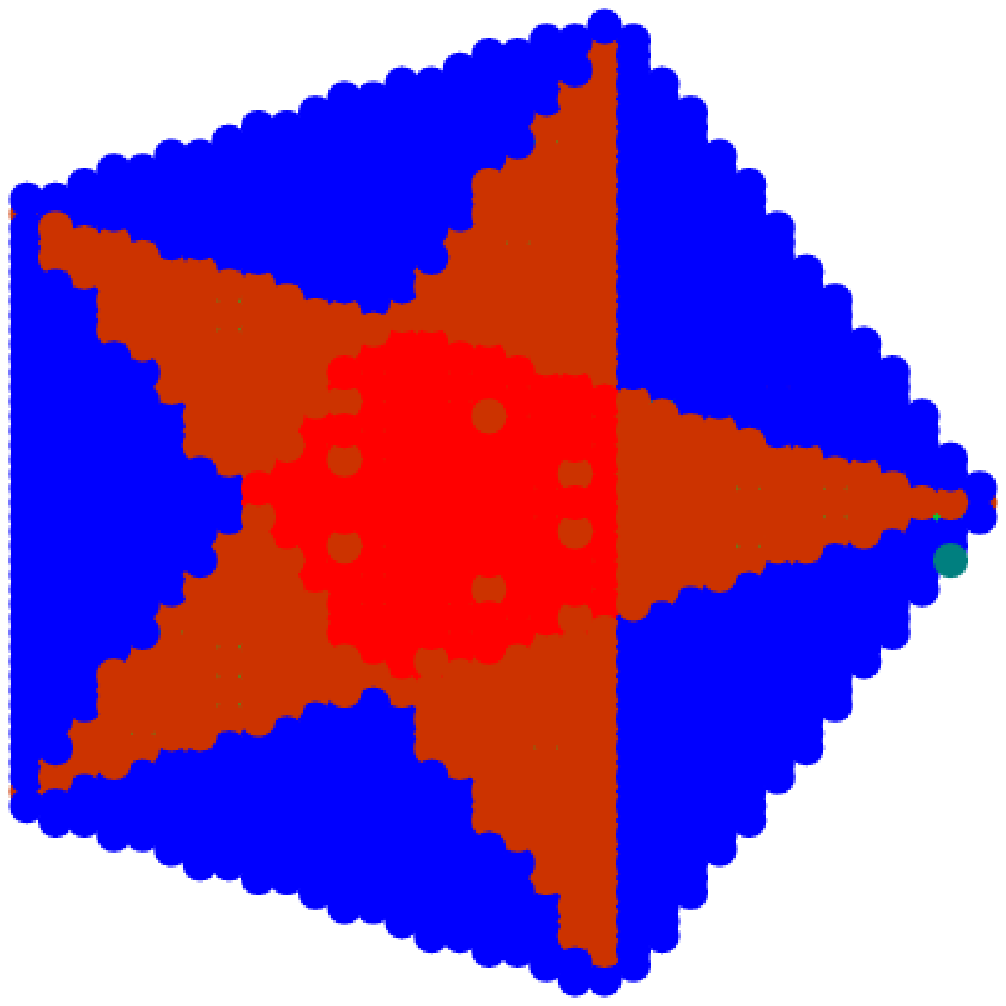}
\includegraphics[scale=0.4]{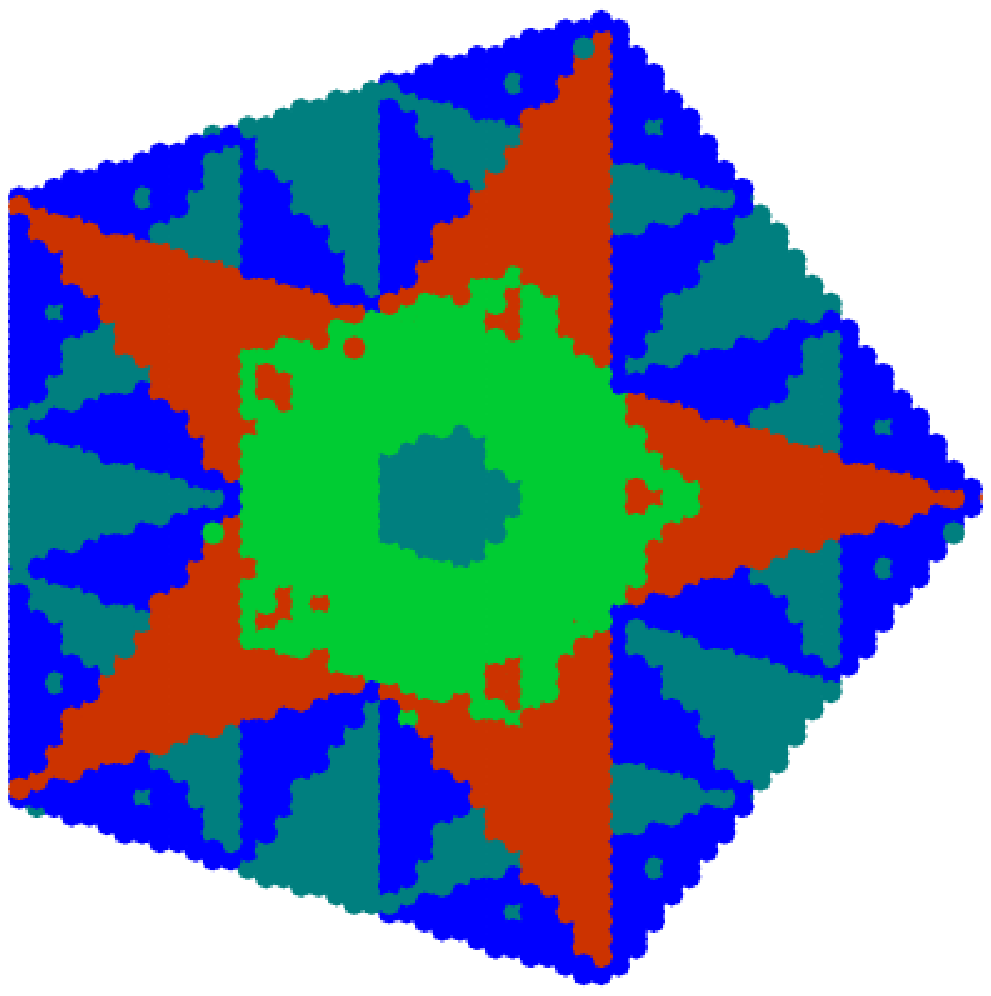}
\vspace{.2cm} \caption{(Color online) Fig.\ref{fig:realspacefig} represented in perpendicular space. Two planes are shown (corresponding to $z_\perp =1,2$) (magnetization values : red (small), green (intermediate) blue (highest))} \label{fig:perpfigmag}
\end{center}
\end{figure}

We show, in Fig.\ref{fig:perpfigmag}, two sets of projections onto perpendicular space of the vertices of the Taylor approximants. The corresponding onsite magnetizations, as calculated for the tiling in real physical space, are colored according to their values as previously, in Fig.\ref{fig:realspacefig}. The figures show that, as expected, the domain corresponding to each coordination number has a distinct color. The colors are not absolutely uniform since no two sites are identical. Self-similar patterns can be investigated using this type of representation, however our system sizes are not large enough to enable a quantitative analysis of self-similarity in the ground state.
Fig.\ref{fig:perpfigmag}a shows the domain corresponding to the F sites, as a central star-shaped region, which has the lowest $m_{si}$ values. The S sites project into a different star-shaped domain shown in Fig.~\ref{fig:perpfigmag}b, and can be seen to have a bigger value of the staggered local magnetization.
We mentioned that the Penrose tiling is invariant under the inflation transformation where edge lengths are expanded by a factor $\tau$.  If there is any self similarity of the ground state under an inflation operation, it can be perceived on the perpendicular space magnetization map as pairs of similar patterns in regions which are related via an inflation transformation. 

\section{Predictions using the Heisenberg star model.}

We first consider the effects of variation of the number of nearest neighbors using a simple Heisenberg star model, as first outlined in \cite{ajrmsw}. A central spin $\bf{S}_0$ is coupled to $z$ neighbors $\bf{S}_j$. The Hamiltonian $H = \sum_{j=1}^z J {\bf S}_0.{\bf S}_j$, can be expanded in boson operators $a$ (describing the center site spin fluctuations) and $b_j$, as described in \cite{ajrmsw}. The corresponding magnetizations are respectively $m_0 = S - \delta m_0$ and $m_j=S-\delta m_j$ where (for $z>1$)

\begin{eqnarray}
\label{eq:mstar} \delta m_0&=&1/(z-1) \\ \nonumber \delta m_j&=&
1/z(z-1)
\end{eqnarray}
Quantum fluctuations of the center spin are thus larger than those of the outer spins. This is a consequence of the fact that in such clusters, the classical term, which creates an onsite potential $V_0 \propto z$, thereby discouraging boson formation on the center site, is dominated by the
transverse terms. These terms in $a^\dag b_j^\dag$ create and annihilate boson pairs on each link, leading to quantum fluctuations being greater for sites with more neighbors.

This cluster model is a first approximation. In
order to explain the multiple peaks in the values of $m_{si}$ seen numerically
for a given $z$, we must take into account longer range structural details.

\begin{figure}[t] \begin{center}
\includegraphics[width=4cm]{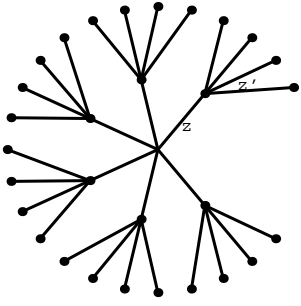}
\caption{\label{fig:hstar} Two-tier Heisenberg star
}
\end{center}\end{figure}

We consider therefore the two tier Heisenberg star shown in Fig.\ref{fig:hstar}, as described in \cite{brep2007}. The Hamiltonian of this cluster is linearized after introducing Holstein-Primakoff operators $a_0, a_i$, ($i=1,...,zz'$) and $b_j$, ($j=1,..,z$) The resulting expression for the center site magnetization is 
\begin{eqnarray}
m_s(z,z') = \frac{1}{2} -\frac{ zf_1^2(z,z')}{ f_2^2(z,z')- zf_1^2(z,z') - 4z'}, \label{hstarmag}
\end{eqnarray}
where $f_{1(2)}= -z' \pm (2-z+\sqrt{4-4z +(z+z')^2})$. 

\begin{figure}[t] \begin{center}
\includegraphics[width=7cm]{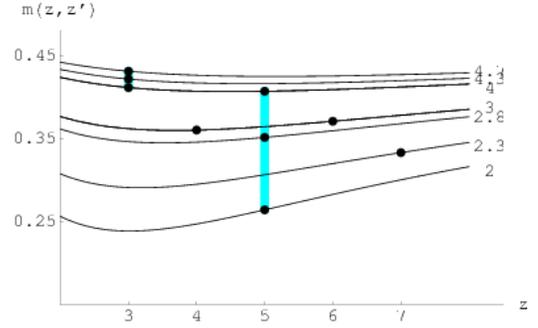}
\caption{\label{fig:predict} $m_s(z,z')$ plotted as a function of $z$ (see text) for selected values of $z'$ (as given in Table 1).}
\end{center}\end{figure}

The main new feature of $m_s(z,z')$ is its non-monotonicity. $m$ is shown plotted against $z$ for various values of $z'$ in Fig.\ref{fig:predict}. For each of the curves of fixed $z'$, a shallow minimum is seen to occur for values $z \sim 1+z'$, that is, when the coordination numbers of the center site and the nearest neighbors are matched.

Returning to the Penrose tiling and approximants thereof, 
we use the number of next nearest neighbor bonds $n(z)$ to classify sites of a given $z$. The
$z=3$ sites can be classified in three subgroups,
depending on $z'$, as shown in Table 1. 
Fig.\ref{fig:distfn} shows the values of $m_s(z,z')$ obtained for the Penrose tiling sites. 
To take the case of $z=3$, three values 
corresponding to three main local configurations are found, to be compared with the three principal peaks of the red curve in Fig.\ref{fig:distfn}. The substructures
arise due to differences in the third nearest and further neighbor
configurations.

The number of next nearest neighbors also serves to distinguish between the different $z=5$ sites, of which there are three main types. There
are two varieties of five-fold symmetric sites: the football-shaped
clusters (F) and the star-shaped clusters (S). The former have 
$z'=2$, while the latter have $z'=4$. The $F$ sites thus have the smallest onsite magnetizations and the S-sites, on the contrary, have the largest onsite magnetization. The remaining (most
frequently occurring) $z=5$ sites which do not have a local five-fold symmetry
 have intermediate values of $z'$.

\begin{center}
\begin{tabular}{|c|c|c|}
\hline
$z$ & $z'$ (corresponding frequency)& $m_s(z,z')$ \\
\hline
3 & 4 (31\%),4.33 (27\%), 4.67 (42\%) & 0.41, 0.42, 0.43 \\
\hline
4 & 3 (100\%) & 0.36 \\
\hline
5 & 2 (14\%),2.4 - 3.2 (81\%), 4 (5\%) & 0.26, 0.35, 0.41\\
\hline
6 & 3 (100\%) & 0.37  \\
\hline
7 & 2.3 (100\%) & 0.33 \\
\hline
\end{tabular}
\vskip 0.5cm \small{Table 1. Values of $z'$ (and their frequencies)
 for each coordination $z$ and the predicted values of $m(z,z')$ using Eq.\ref{hstarmag}}
\end{center}

\subsection{Fourier transformation}
\begin{figure}[t] \begin{center}
\includegraphics[width=5cm]{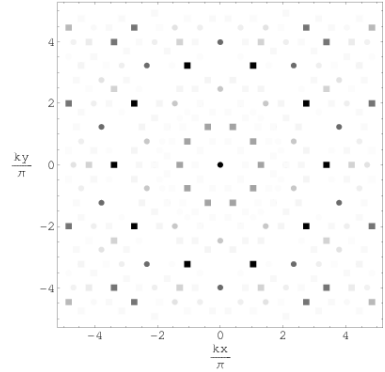}
\caption{\label{fig:fourier}
Intensity plot of the static longitudinal magnetic structure factor $S^{\parallel}({\mathbf k})$ for the $N= 4414$ Taylor approximant. The highest intensity peaks are shown in the figure at positions that are indicated by circles(squares) for the magnetic (nonmagnetic) structures respectively. The relative intensity is denoted by a linear gray scale ranging between zero (white) and maximum intensity (black).
}
\end{center}\end{figure}

Fourier transforms of the original structure, as well as that of the antiferromagnetic ground states are shown in \ref{fig:fourier}. The highest intensity peaks are shown in the figure at positions that are indicated by circles(squares) for the magnetic (nonmagnetic) structures respectively. Peaks can be indexed using four indices which are integers for non-magnetic peaks and integer or half-integer for the magnetic peaks. As in the case of the octagonal tiling \cite{wess2} the half integer indices arise  due to the doubling in size of the antiferromagnetic unit cell in five dimensions. This, coupled with the extinctions in the structure factor rule due to the multiplicity of each unit cell, then leads \cite{lifshitz} to the observed ``shifting" or ``displacement" of the magnetic peaks with respect to the nonmagnetic ones.

\section{Conclusions}
We have presented a detailed analysis of energies and wavefunctions of magnon eigenmodes found by numerical diagonalization of the Heisenberg model on approximants of the Penrose tiling in the linearized spin wave approximation. 
A linear dispersion is found at low energies, as in other bipartite antiferromagnets, despite the lack of translational invariance. This is to be expected in the long wavelength limit, where structural details do not play a role, as known in related studies of low energy vibrational modes in quasicrystals or in glasses. The effective spin wave velocity as obtained from the low energy spectrum was found and compared with that in the square lattice and the octagonal tiling. 

The spatial dependence of eigenmodes has been investigated. It is shown that in different ranges of energies, states have their support on sites of a different given coordination number. At low energies, in contrast, all sites participate, and the resulting eigenstates are relatively extended, as seen by the absolute values and the size dependence of the participation ratio. A few low energy extended eigenstates are analysed for their multifractal scaling properties, including the similarity dimension and information dimensions. In general, our studies indicate that the dimensionality of the states diminishes progressively as energy increases.

We find the ground state energy of this antiferromagnet, and give the distribution of the local magnetizations as a function of site coordination number. Perpendicular space projections are shown to demonstrate the simplicity of the ground state in this representation. The structure factor of the magnetic state is found and compared with that of the nonmagnetic state.

A simple analytical model, the two-tier Heisenberg star, is presented to explain our results. The role of next neighbors is shown to be nontrivial, and explains the $z$ dependence of the onsite staggered magnetizations. 

\acknowledgments
We would like to thank M. Duneau and S. Wessel for many useful discussions. A.Sz. was supported by the European Commission, through a Marie Curie Foundation contract, MEST CT 2004-51-4307, during the course of this work. 

\appendix
\section{Obtaining the Penrose tiling and its approximants by projection}
\subsection{The perfect Penrose tiling.}
The perfect Penrose tiling can be obtained by the projection onto the $xy$ plane of selected vertices of
a five dimensional (5D) cubic lattice. A vertex, designated by the five-dimensional vector
${n_1,n_2,....,n5}$, is selected for projection if its projection in the three dimensional ``perpendicular
space" belongs in the region $W$ (called the ``selection window").
The projection matrices $M_\parallel$ and $M_\perp$ give the real space coordinates $x_\parallel,y_\parallel$ and perpendicular space coordinates $x_\perp,y_\perp,z_\perp$ respectively are

\begin{eqnarray}
M_\parallel = \sqrt{\frac{2}{5}}\left(
\begin{array}{rrrrr}
1&\cos \theta&\cos 2\theta& \cos 2\theta&\cos \theta\\
0&\sin \theta&\sin 2\theta &-\sin 2\theta & -\sin \theta\\
\end{array}
\right)
\end{eqnarray}

\begin{eqnarray}
M_\perp = \sqrt{\frac{2}{5}}\left(
\begin{array}{rrrrr}
1&\cos 2\theta&\cos \theta& \cos \theta&\cos é\theta\\
0&\sin 2\theta&-\sin \theta &\sin \theta & -\sin 2\theta\\
\frac{1}{\sqrt{2}}&\frac{1}{\sqrt{2}}&\frac{1}{\sqrt{2}}&\frac{1}{\sqrt{2}}&\frac{1}{\sqrt{2}}\\
\end{array}
\right)
\end{eqnarray}

where $\theta = 2\pi/5$.

\begin{figure}[t] \begin{center}
\includegraphics[width=5cm]{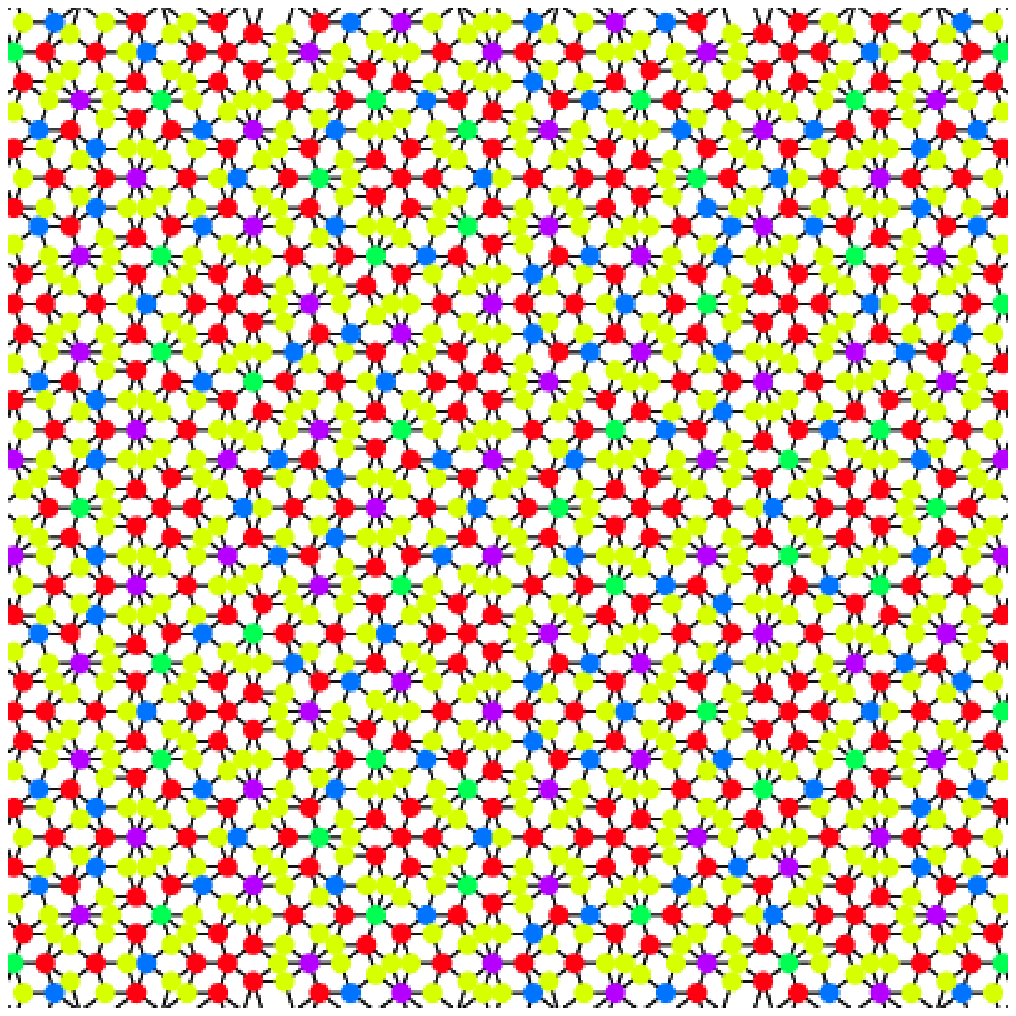}
\includegraphics[width=8cm]{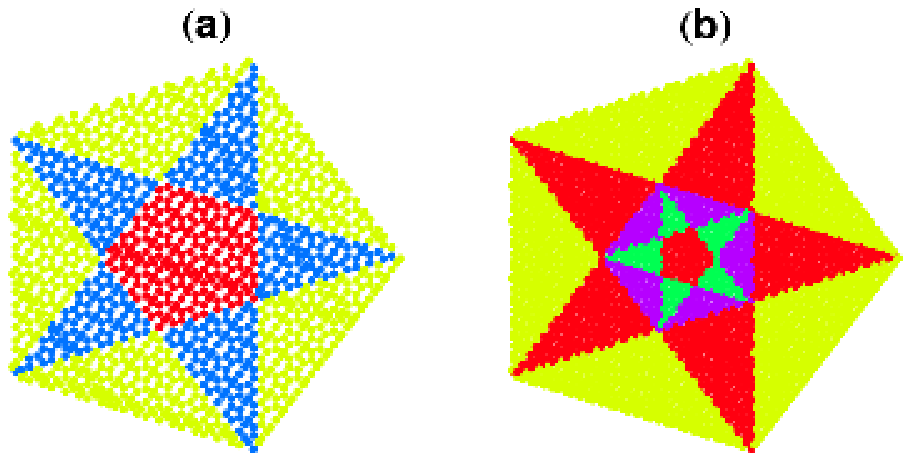}
\caption{\label{fig:realspace} (Color online) (upper) Portion of the tiling showing vertices colored differently, according to coordination number $z$ (lower) The same tiling after projection into perpendicular space (a) the plane $z_\perp =1$, b) the plane $z_\perp =2$ )}
\end{center}\end{figure}

For a vertex to be selected, its perpendicular coordinates must fall within the rhombic icosahedron $W$ which is the projection of the 5D unit cube. In view of the expression for $M_\perp$ given above, $W$ decomposes into four subdomains $W_i$, corresponding to $z_\perp =1,2,3,$ or $4$ mod(5). The acceptance window $W_i$ in each of these four planes in the perpendicular space is a pentagon. Two such pentagons are shown in Fig.\ref{fig:realspace} (corresponding to $z_\perp = 1$ and $2$), the other two being the same upto an inversion.

The parity of the vertex as given by $\sum n_i$ determines the sublattice to which it belongs. Thus, the points that project into the planes $z_\perp =1,3$ correspond, say, to sublattice A , while the planes$z_\perp =2,4$  correspond to sublattice B. In the infinite tiling, the two sublattices are equivalent, and the ``even" and ``odd" windows are the same upto an inversion.

Finite samples of arbitrarily large size can be readily generated using this selection criterion, by considering large enough volumes of the 5D cubic lattices.

Turning to the question of the local environments, we show in Fig.\ref{fig:realspace} a portion of the Penrose tiling, with vertices of different $z$ colored differently. When these vertices are represented in perpendicular space, the colored domains are clearly seen corresponding to each value of $z$. We have seen that there are three types of $z=5$ sites, including the F and S subsets. The F sites project onto the central pentagonal regions in the planes $z_\perp =1$ and 4, while the S sites project into the central pentagons in the planes  $z_\perp =2$ and 3.

\subsection{The Taylor approximants.}
\begin{figure}[t] \begin{center}
\includegraphics[width=4cm]{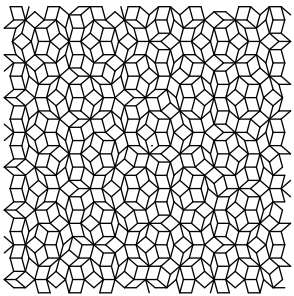}
\caption{\label{fig:taylor3} Taylor $\tau^3$ approximant (N=644 sites)}
\end{center}\end{figure}

To obtain periodic approximants, the physical plane has to be tilted so that it has a rational orientation (that can be chosen arbitrarily close to the original irrational orientation). There will then be two lattice vectors $\vec{A}_1$ and $\vec{A}_2$ of the 5D lattice, whose projections give the periodic lengths in the physical plane.
This is achieved in practice by using an oblique projection $\tilde{M}_\perp$ appropriately defined depending on the choice made for $\vec{A}_1$ and $\vec{A}_2$. The resulting selection window $\tilde{W}$ will be a deformed rhombic icosahedron. Details of the construction of the projection operators for the so-called Taylor approximant can be found in \cite{dunaudier}. This method can be used to obtain bigger systems, and we have considered six approximants, that we labelled Taylor, Taylor $\tau$, ..., upto Taylor $\tau^6$ (containing 11556 sites). (Note: the original Taylor approximant, consisting of 36 sites is too small for use in our analysis).
As an example, in the case of our Taylor $\tau^3$ approximant  (N=644 sites), we have

\begin{eqnarray}
\vec{A}_1 &=& \{10, 3, -8, -8, 3\} \\
\vec{A}_2 &=& \{0, 8, 5, -5, -8\}
\end{eqnarray}

with the resulting rectangular shape of sides $L_x = 24.7984$ and $L_y = 21.0948$ (see Fig.\ref{fig:taylor3}).

Note: In our calculations, we have used systems having sublattices A and B of the same size. This was achieved by shifting the selection windows in perpendicular space, until the number of A and B sublattice points selected are equal.

\end{document}